\documentstyle[11pt]{article}

\let\a=\alpha \let\b=\beta   
    
  \let\n=\nu

\def\nn{\nonumber} \def\bd{\begin{document}} \def\ed{\end{document}}
\def\ds{\documentstyle} \let\fr=\frac \let\bl=\bigl \let\br=\bigr
\let\Br=\Bigr \let\Bl=\Bigl 
\let\bm=\bibitem
\let\na=\nabla
\let\pa=\partial \let\ov=\overline 
\newcommand{\be}{\begin{equation}} 
\newcommand{\ee}{\end{equation}} 
\def\ba{\begin{array}}
\def\ea{\end{array}}
\def\ft#1#2{{\textstyle{{\scriptstyle #1}\over {\scriptstyle #2}}}}
\def\fft#1#2{{#1 \over #2}}
\def\del{\partial}
\def\vp{\varphi}
\def\sst#1{{\scriptscriptstyle #1}}
\def\oneone{\rlap 1\mkern4mu{\rm l}}
\def\td{\tilde}
\def\wtd{\widetilde}
\def\ie{\rm i.e.\ }
\def\dalemb#1#2{{\vbox{\hrule height .#2pt
        \hbox{\vrule width.#2pt height#1pt \kern#1pt
                \vrule width.#2pt}
        \hrule height.#2pt}}}
\def\square{\mathord{\dalemb{6.8}{7}\hbox{\hskip1pt}}}
\newcommand{\ho}[1]{$\, ^{#1}$}
\newcommand{\hoch}[1]{$\, ^{#1}$}
\newcommand{\bea}{\begin{eqnarray}} 
\newcommand{\eea}{\end{eqnarray}} 
\newcommand{\ra}{\rightarrow}
\newcommand{\lra}{\longrightarrow}
\newcommand{\Lra}{\Leftrightarrow}
\newcommand{\ap}{\alpha^\prime}
\newcommand{\bp}{\tilde \beta^\prime}
\newcommand{\tr}{{\rm tr} }
\newcommand{\Tr}{{\rm Tr} } 
\def\0{{\sst{(0)}}}
\def\1{{\sst{(1)}}}
\def\2{{\sst{(2)}}}
\def\3{{\sst{(3)}}}
\def\4{{\sst{(4)}}}
\def\5{{\sst{(5)}}}
\def\6{{\sst{(6)}}}
\def\7{{\sst{(7)}}}
\def\8{{\sst{(8)}}}
\def\n{{\sst{(n)}}}
\def\tV{\widetilde V}
\def\tW{\widetilde W}
\def\tH{\widetilde H}
\def\tE{\widetilde E}
\def\tF{\widetilde F}
\def\tA{\widetilde A}
\def\im{{{\rm i}}}
\def\tY{{{\wtd Y}}}
\def\ep{{\epsilon}}
\def\vep{{\varepsilon}}

\newcommand{\NP}{Nucl. Phys. }
\newcommand{\tamphys}{\it Center for Theoretical Physics,
Texas A\&M University, College Station, TX 77843}
\newcommand{\upenn}{\it Dept. of Phys. and Astro.,
University of Pennsylvania,
Philadelphia, PA 19104}

\newcommand{\auth}{M. Cveti\v{c}\hoch{\dagger1}, H. L\"u\hoch{\dagger1},
C.N. Pope\hoch{\ddagger2} and A. Sadrzadeh\hoch{\ddagger}}

\begin{document}
\begin{flushright}
\hfill{CTP TAMU-04/00 \\
UPR/876-T \\
February 2000}\\
\hfill{\bf hep-th/0002056}\\
\end{flushright}

\vspace{10pt}

\begin{center}
{\large {\bf Consistency of Kaluza-Klein Sphere Reductions of
Symmetric Potentials}}

\vspace{20pt}

\auth

\vspace{10pt}
{\hoch{\dagger}\upenn}

\vspace{10pt}
{\hoch{\ddagger}\tamphys}

\vspace{30pt}

\underline{ABSTRACT}
\end{center}

   In a recent paper, the complete (non-linear) Kaluza-Klein Ansatz
for the consistent embedding of certain scalar plus gravity subsectors
of gauged maximal supergravity in $D=4$, 5 and 7 was presented, in
terms of sphere reductions from $D=11$ or type IIB supergravity.  The
scalar fields included in the truncations were the diagonal fields in
the $SL(N,R)/SO(N)$ scalar submanifolds of the full scalar sectors of
the corresponding maximal supergravities, with $N=8$, 6 and 5.  The
embeddings were used for obtaining an interpretation of extremal
$D=4$, 5 or 7 AdS domain walls in terms of distributed M-branes or
D-branes in the higher dimensions.  Although strong supporting
evidence for the correctness of the embedding Ans\"atze was presented,
a full proof of the consistency was not given.  Here, we complete the
proof, by showing explicitly that the full set of higher-dimensional
equations of motion are satisfied if and only if the lower-dimensional
fields satisfy the relevant scalar plus gravity equations.

{\vfill\leftline{}\vfill
\vskip 10pt \footnoterule {\footnotesize \hoch{1} Research supported
in part by DOE grant DOE-FG02-95ER40893
\vskip  -12pt} \vskip   14pt
{\footnotesize
        \hoch{2}        Research supported in part by DOE
grant DOE-FG03-95ER40917 \vskip -12pt}  \vskip  14pt
}

\pagebreak
\setcounter{page}{1}

\section{Introduction}

    One of the more intriguing outcomes of recent work on the AdS/CFT
correspondence has been a renewed effort to understand how the
lower-dimensional gauged supergravities arise as Kaluza-Klein sphere
reductions from $D=11$ or type IIB supergravity.  It was long ago
demonstrated how the reductions work at the linearised level, but few
complete non-linear results existed.  A proof of the consistency of
the $S^7$ reduction from $D=11$ was presented, although the
Kaluza-Klein Ansatz for the field-strength sector was not
fully explicit \cite{deWitnicolai}.  It was generally assumed that the
other cases, namely the $S^4$ reduction of $D=11$, and the $S^5$
reduction of type IIB, would be consistent too, but until recently no
results for these cases had been obtained.  In recent work, a fully
explicit reduction Ansatz for the $SO(5)$-gauged $N=4$ $D=7$ case has
been obtained \cite{nvv0,nvv}.\footnote{The complete bosonic reduction
Ansatz for another case, namely the local $S^4$ reduction of massive
type IIA supergravity to $SU(2)$ gauged $N=2$ supergravity in $D=6$
has also been obtained \cite{d6gauge}.}  Explicit results have also
been obtained for various truncations of the full maximal
supergravities.  These include truncations to the maximal abelian
subgroups $U(1)^4$, $U(1)^3$ and $U(1)^2$ in $D=4$, 5 and 7
\cite{ten}; the truncation to $SU(2)$-gauged $N=2$ in $D=7$
\cite{d7gauge}; to $SU(2)\times U(1)$ gauged $N=4$ in $D=5$
\cite{d5gauge}; and to $SO(4)$ gauged $N=4$ in $D=4$ \cite{d4gauge}.
For many purposes, if the fields that participate in the
lower-dimensional solutions of interest lie within these truncated
subsectors, the truncated reduction is much easier to use, since it is
usually much simpler than the full maximal result.\footnote{We
emphasise that in this discussion we are considering only the
``remarkable'' Kaluza-Klein sphere reductions, where there is no
group-theoretic understanding for why the consistency is achievable.
In particular, some of the scalar fields parameterise inhomogeneous
distortions of the sphere.  These contrast with, for example, torus
reductions, where the consistency of the truncation to the massless
sector is guaranteed by group theory.}

   Another truncation that allows for relatively simple although still
non-trivial sphere reductions is where one retains only the metric
and a certain subset of the scalar fields of the lower-dimensional
gauged supergravity; one can keep just certain scalars
contained in the $SL(N,R)/SO(N)$ subset of the full scalar coset
manifold, which can be described by a symmetric tensor $T_{ij}$.  
In $D=4$, 5 and 7 these subsets correspond to $N=8$, 6 and
5 respectively.  Specifically, one can consistently truncate to the diagonal
scalars, $T_{ij}= X_i\, \delta_{ij}$, where $\prod_i X_i=1$.  Thus
there are 7, 5 and 4 independent scalars in the $D=4$, 5 and 7 cases.
As shown in \cite{dist}, where the reduction Ans\"atze for these scalar
subsectors were presented, one can actually discuss all cases where
the lower dimension $D$ is related to $N$ by
\be
N= \fft{4(D-2)}{D-3}\label{ndrel}\,,
\ee
corresponding to supersymmetric higher-dimensional theories, in a
uniform way.  The only integer possibilities are $(D,N)= (4,8)$,
$(5,6)$ and $(7,5)$, as listed above.  (Some proposals for other
scalar truncations were presented recently in \cite{nava}.)  In
\cite{dist}, extremal AdS domain wall solutions in these dimensions
were derived, with the general set of $(N-1)$ independent charge
parameters.  By using the reduction Ans\"atze to oxidise the solutions
to the higher dimensions, it was shown how they can be interpreted as
continuous distributions of M-branes or D-branes \cite{dist}.
(Various special cases were obtained also in
\cite{klt,fgpw,BS,BS2,BBS}.)

   Certain consistency checks for the reduction Ans\"atze presented in
\cite{dist} were conducted there, but a full demonstration of the
consistency was not given.  Here, we complete the argument by checking
all the higher-dimensional equations of motion, and verifying that
indeed they are satisfied by the Ans\"atze of \cite{dist}, if and only
if the lower-dimensional equations of motion are 
satisfied.\footnote{Note that substituting into the higher-dimensional
Lagrangian and integrating out the sphere directions could never, {\it
per se}, yield a proof of consistency.}  Of course
these calculations would be subsumed by complete demonstrations of the
consistency of the maximal supergravity reductions in $D=4$, 5 and 7.
Such a complete proof exists for $D=7$ \cite{nvv0,nvv}, and implicitly for
$D=4$ \cite{deWitnicolai}, but not yet for $D=5$.  Thus the results
presented here provide new and independent evidence for the
conjectured consistency in all the cases.  

\section{The Scalar Theories, and the Reduction Ans\"atze}

    The truncated lower-dimensional gravity plus scalar theory is
described by the following Lagrangian in $D$ dimensions \cite{dist}:
\be
e^{-1}\, {\cal L}_D = R -\ft12 (\del\vec\varphi)^2 - V\,,\label{ddlag}
\ee
where the potential $V$ is given by
\be
V = -\ft12 g^2\, \Big( (\sum_{i=1}^N
 X_i)^2 - 2 \sum_{i=1}^N X_i^2 \Big)\,.\label{ddpot}
\ee
(In $D=4$, 5 and 7, we shall have $N=8$, 6 and 5 respectively.)  The
$N$ quantities $X_i$ are parameterised in terms of $(N-1)$ independent
dilatonic scalars $\vec\varphi$ as follows:
\be
X_i = e^{-\fft12 \vec b_i\cdot\vec\varphi}\,,\label{ddxdef}
\ee
where the $\vec b_i$ satisfy
\be
\vec b_i\cdot\vec b_j = 8\delta_{ij} - \fft{8}{N}\,,\qquad
\sum_i \vec b_i=0\,,\qquad
(\vec u\cdot\vec b_i)\, \vec b_i = 8 \vec u\,,\label{dotprod}
\ee
The middle equation here expresses the fact that the $N$ quantities
$X_i$ are subject to the condition
\be
\prod_{i=1}^N X_i = 1\,.\label{prodcon}
\ee
The last equation in (\ref{dotprod}) allows us to express the dilatons
$\vec\varphi$ in terms of the $X_i$:
\be
\vec\varphi = -\ft14 \sum_i \vec b_i\, \log X_i\,.\label{phiexp}
\ee

      The equations of motion for the scalar fields, following from
(\ref{ddlag}), are
\be
\square \vec\varphi = \fft{\del V}{\del\vec\varphi}\,.\label{ddeom}
\ee
 From (\ref{ddxdef}) it follows that $\del X_i/\del\vec\varphi =
-\ft12 \vec b_i\, X_i$, and hence the equations of motion
(\ref{ddeom}) become
\be
\square\vec\varphi = \ft12 g^2\, \sum_i \vec b_i
\Big( X_i\, \sum_j X_j  - 2  X_i^2\Big)\,.\label{phieq}
\ee
Note that we can also write the scalar equations of motion as
\be
\square\log X_i = 2 g^2\, \Big( 2 X_i^2 - X_i\, \sum_j X_j - \ft2{N}
\sum_k X_k^2 + \ft1{N} (\sum_j X_j)^2 \Big)\,.\label{xeq0}
\ee
The Einstein equation following from (\ref{ddlag}) is
\be
R_{\mu\nu} = \ft14 X_i^{-2}\, \del_\mu X_i\, \del_\nu X_i  +
\fft1{D-2}\, V\, g_{\mu\nu}\,.\label{ddeinst}
\ee

   The Kaluza-Klein sphere reduction Ans\"atze for obtaining these
theories from the higher dimension were presented in \cite{dist}, and are
as follows:
\bea
d\hat s^2 &=& \Delta^{\fft2{D-1}} \, ds_D^2 + \fft1{g^2}\,
\Delta^{-\fft{D-3}{D-1}}\, \sum_i X_i^{-1}\, d\mu_i^2\,,\nn\\
\hat F &=& g\, \sum_i(2 X_i^2\, \mu_i^2 - \Delta\, X_i)\,
\ep_{\sst{(D)}}
+\fft1{2g}\, \sum_i X_i^{-1} 
\, {* dX_i}\wedge d(\mu_i^2)\,,\label{dgenansatz}
\eea
where
\be
\Delta = \sum_i X_i\, \mu_i^2\,,
\ee
and the $\mu_i$ are a set of $N$ ``direction cosines'' that satisfy
the constraint
\be
\sum_i \mu_i^2 = 1\,.\label{mucon}
\ee
In (\ref{dgenansatz}), $\ep_{\sst{(D)}}$
denotes the volume form of the $D$-dimensional metric $ds_D^2$.  Note
that if all the scalars $X_i$ are trivial ($X_i=1$), the internal part
of the metric becomes $\sum_i d\mu_i^2$, which is the metric on the
unit $(N-1)$-sphere.

   The $D$-form field strength $\hat F$ in
(\ref{dgenansatz}) is the 4-form of eleven-dimensional supergravity
for the case $D=4$, the Hodge dual of this 4-form for the case $D=7$,
and it is the self-dual 5-form of the type IIB theory when $D=5$.
Note that in each case, given the nature of the Ansatz, the relevant
Bianchi identity and field equation for $\hat F$ are
simply
\be
d\hat F = 0\,,\qquad d{\hat*  \hat F} =0\,.
\ee

\section{The Consistency of the Reduction}

     It was shown in \cite{dist} that the $D$-form field-strength
Ansatz in (\ref{dgenansatz}) satisfies the Bianchi identity $d\hat
F=0$, provided that the scalar fields $X_i$ satisfy precisely the
lower-dimensional equations of motion (\ref{xeq0}).  This calculation
is a straightforward one, and we shall not repeat it here.  It is
harder to show that $\hat F$ satisfies the field
equation $d{\hat *\hat F} =0$, because this involves taking a Hodge dual
of the field strength $\hat F$.  This is what we shall now address.

\subsection{The Field Equation for $\hat F$}

    The complication here is that the $(N-1)$-sphere is being
coordinatised by $N$ quantities $\mu_i$ subject to the constraint
(\ref{mucon}).  It seems that the best way to proceed is to eliminate
one of the $\mu_i$ in favour of the others, using (\ref{mucon}).  To
that end, we split the $\mu_i$ as $\mu_i=(\mu_\a,\mu_0)$, and solve
for $\mu_0$ in terms of the $\mu_\a$.

      If we consider first the metric 
\be
ds^2=  \sum_i X_i^{-1}\, d\mu_i^2\,,
\ee
then in terms of the $\mu_\a$ we can write it as $ds^2= g_{\a\b}\,
d\mu_\a\, d\mu_\b$, where
\be
g_{\a\b} = X_\a^{-1}\, \delta_{\a\b} + \fft1{X_0\, \mu_0^2}\, \mu_\a\,
\mu_\b\,.\label{downmet}
\ee
(We have, of course, used the identity
\be
d\mu_0 = -\fft{\mu_\a}{\mu_0}\, d\mu_\a\,,
\ee
which follows from (\ref{mucon}).)
   
    It is straightforward to invert the metric $g_{\a\b}$ given in
(\ref{downmet}).  The result is
\be
g^{\a\b} = X_\a\, \delta_{\a\b} - \Delta^{-1}\, \mu_\a\mu_\b\, X_\a\, 
X_\b\,.\label{upmet}
\ee
It is also easy to establish that
\be
\det(g_{\a\b}) = \fft{\Delta}{\mu_0^2}\,.
\ee
Note that it follows from the metric Ansatz in (\ref{dgenansatz})
that the determinant of the higher-dimensional metric $d\hat s^2$ is
given by
\be
\det(\hat g) = \fft{\Delta^{\fft{4}{D-1}}}{g^{2N-2}\, \mu_0^2}\,
\det(g_D)\,,
\ee
where $g_D$ denotes the $D$-dimensional spacetime metric $ds_D^2$ and $g$ 
in the denominator is just the gauge coupling constant (not to be confused with
the determinant of the higher-dimensional metric $\hat g$ or the one 
for the lower dimension, $g_D$.) 

    Now let us look at the field-strength Ansatz.  We shall use the
convention that $\vep_{M_1\cdots M_D}$ always means the tensor
density, which is the pure numbers $\pm 1, 0$.  So the Ansatz for
$\hat F$ in (\ref{dgenansatz}) is
\bea
\hat F_{\nu_1\cdots \nu_D} &=& g U \sqrt{-g_D}\, \vep_{\nu_1\cdots
\nu_D}\,,\nn\\
\hat F_{\nu_1\cdots \nu_{D-1}\a} &=& \fft{1}{g} \sqrt{-g_D}\, 
\vep_{\nu_1\cdots \nu_{D-1} \rho}\, g_D^{\rho\sigma}\,
(X_\a^{-1}\, \del_\sigma X_\a - X_0^{-1}\, \del_\sigma X_0)\,
\mu_\a\,,\label{fans}
\eea
where
\be
U \equiv  \sum_i(2 X_i^2\, \mu_i^2 - \Delta\, X_i)\,,
\ee
and $g_{D}$ denotes the $D$-dimensional spacetime metric $ds_D^2$.

   We can now calculate the upper-index components of $\hat F$.  In
fact, what we really need is these components multiplied by
$\sqrt{\hat g}$.  From the results above we find
\bea
\sqrt{-\hat g}\, \hat F^{\nu_1\cdots \nu_D} &=& \fft{U}{g^{N-2} \mu_0\,
\Delta^2}\, \vep^{\nu_1\cdots \nu_D}\,,\nn\\
\sqrt{-\hat g}\, \hat F^{\nu_1\cdots \nu_{D-1}\a}
&=&\fft1{g^{N-2} \mu_0}\, \vep^{\nu_1\cdots\nu_{D-1}\, \sigma}\,  
\del_\sigma\Big( \fft{X_\a\,
\mu_\a}{\Delta} \Big)\,.\label{fup}
\eea
($\vep^{M_1\cdots M_D}$ is the tensor density that takes the values
0, $\pm1$, and is numerically equal to $\vep_{M_1\cdots M_D}$.)

    One can directly verify from these expressions that the field
equation is satisfied, namely that
\be
\del_M\Big( \sqrt{-\hat g}\, \hat F^{N_1\cdots N_{D-1} M} \Big)=0\,.
\ee
However, it is more elegant to do this by using (\ref{fup}) first to
construct the Hodge dual of
$\hat F$ itself.  To do this, we make the following definitions:
\bea
P &\equiv& \fft1{n!}\, \vep_{\a_1\cdots\a_n}\, d\mu_{\a_1}\cdots
d\mu_{\a_n}\,,\nn\\
Q_\a &\equiv & \fft1{(n-1)!}\, \vep_{\a\b_1\cdots \b_{n-1}}\,
d\mu_{\b_1}\cdots d\mu_{b_{n-1}}\,,\nn\\
W &\equiv &  \fft1{n!}\, \vep_{i j_1\cdots j_n}\, \mu_i\, 
d\mu_{j_1}\cdots d\mu_{j_n}\,,\nn\\
Z_i &\equiv & \fft1{(n-1)!}\, \vep_{ij k_1\cdots k_{n-1}}\, \mu_j\,
d\mu_{k_1}\cdots d\mu_{k_{n-1}}\,,
\eea
where $n=N-1$.  Note that what we have done here is to define $P$ and
$Q_\a$ with respect to the reduced set of $n=N-1$ coordinates 
$\mu_\a$, while $W$
and $Z_i$ are defined with respect to the full set of $N$ coordinates
$\mu_i$.  (Some analogous formulae and manipulations are presented also in 
\cite{nvv}.)

    Now, we can establish the following:
\bea
W &=& \fft1{\mu_0}\, P\,,\nn\\
Z_0 &=& \mu_\b\, Q_\b\,,\qquad Z_\a = \fft1{\mu_0}\, (-Q_\a + \mu_\a\,
\mu_\b\, Q_\b)\,,\nn\\
d\mu_\a\wedge Q_\b &=& P\, \delta_{\a\b}\,,\nn\\
d\mu_i\wedge Z_j &=& -(\delta_{ij}- \mu_i\, \mu_j)\, W\,,\nn\\
dQ_\a &=& 0\,,\qquad dW=0\,,\qquad dZ_i = n\, \mu_i\, W\,.
\eea

   From (\ref{fup}), it is evident that we have
\be
{\hat *\hat F}_{\a_1\cdots\a_n} = \fft{U}{g^{N-2} \mu_0\, \Delta^2}\, 
\vep_{\a_1\cdots\a_n}\,,\qquad
{\hat *\hat F}_{\a_1\cdots\a_{n-1}\nu } =-\fft{1}{g^{N-2} \mu_0}\, 
\vep_{\a_1\cdots\a_{n-1}\b}\, \del_\nu \Big( \fft{X_\b\,
\mu_\b}{\Delta} \Big)\,.
\ee
Note that here, and in many other formulae, we are using a
``generalised Einstein summation convention,'' in which any dummy
index that appears two or more times in an expression is understood to
be summed over.  It will always be clear from context whether an index
is a dummy or not.

    After some algebra, we can show from the above definitions and
properties that this can be written as
\be
{\hat * \hat F}=\fft{1}{g^{N-2}} \Big( \fft{U}{\Delta^2}\, W +\del_\nu\big
( \fft{X_i\, \mu_i}{\Delta}\big)\, dx^\nu\wedge Z_i\Big)\,.
\ee
To check that the equation of motion $d{\hat *\hat F}=0$ is
satisfied, we just have to make use of the various lemmata established
above.  Thus we have
\bea
g^{N-2} d{\hat *\hat F} &=& \del_\nu\, \Big(\fft{U}{\Delta^2}\Big) \,
dx^\nu\wedge W  - \del_\nu\Big ( \fft{X_i\, \mu_i}{\Delta}\Big)\, dx^\nu\wedge
dZ_i - \del_{\mu_j}\del_\nu \Big ( \fft{X_i\, \mu_i}{\Delta}\Big)\, dx^\nu
\wedge d\mu_j\wedge Z_i\,,\nn\\
&=&  \del_\nu\, \Big(\fft{U}{\Delta^2}\Big) \,
dx^\nu\wedge W - 
n\,\mu_i\, \del_\nu\Big ( \fft{X_i\, \mu_i }{\Delta}\Big)\, dx^\nu\wedge W
\nn\\
&&+  \del_{\mu_j}\del_\nu \Big ( \fft{X_i\, \mu_i}{\Delta}\Big)\, dx^\nu
\wedge W\, (\delta_{ij} - \mu_i\, \mu_j) \,,\nn\\
&=&  \del_\nu\, \Big(\fft{U}{\Delta^2}\Big) \,
dx^\nu\wedge W + \del_\nu\Big( \delta_{ij}\, \fft{X_i}{\Delta} - 
\fft{2 X_i\, X_j\, \mu_i\, \mu_j}{\Delta^2} \Big) \, dx^\nu
\wedge W\, (\delta_{ij} - \mu_i\, \mu_j)\,,\nn\\
&=&  \del_\nu\, \Big(\fft{U}{\Delta^2}\Big) \,
dx^\nu\wedge W -\del_\nu\, \Big(\fft{U}{\Delta^2}\Big) \,
dx^\nu\wedge W=0\,.
\eea
Note that in various steps above, we have made use of the fact
that the $\mu_i$ can be taken freely inside the $\del_\nu$ derivative,
and that therefore, for instance, a term like $\mu_i\, \del_\nu( X_i\,
\mu_i/\Delta)$ is equal to $\del_\nu(X_i\, \mu_i^2/\Delta)$, which is
therefore zero since $X_i\, \mu_i^2 = \Delta$.  
This completes the checking of the consistency of the
higher-dimensional field equation for $\hat F$.

\subsection{The Einstein Equation}

\subsubsection{Calculation of the Ricci Tensor}

    To check the various components of the higher-dimensional Einstein
equation, we first calculate the curvature tensor for the metric
Ansatz.  From now on, since no generality is lost, we set the gauge
coupling $g$ equal to 1 for simplicity.  The metric can be written as
\be
d\hat s^2 = \Delta^a\, ds_D^2 + \Delta^{-b}\, \sum_i X_i^{-1}\,
d\mu_i^2\,,
\ee
where
\be
a= \fft2{D-1}\,,\qquad b= \fft{D-3}{D-1}\,.
\ee

   From this, we find that the affine connection
$\hat \Gamma^M{}_{NP} = \ft12 \hat g^{MQ}\, (\del_N\, g_{QP} +
\del_P\, g_{QN} - \del_Q\, g_{NP})$ is given by
\bea
\hat\Gamma^\mu{}_{\nu\rho} &=&  \Gamma^\mu{}_{\nu\rho} + \ft12 a\,
\Delta^{-1}\, (\delta^\mu_\rho\, \del_\nu\, \Delta + \delta^\mu_\nu\,
\del_\rho\, \Delta -g_{\nu\rho}\,  \del^\mu\, \Delta)\,,\nn\\
\hat\Gamma^\mu{}_{\nu\a} &=& \ft12 a\, \Delta^{-1}\, \delta^\mu_\nu\,
\del_\a\, \Delta\,,\nn\\
\hat \Gamma^\a{}_{\mu\nu} &=& -\ft12 a\, g_{\mu\nu}\, \del^\a\,
\Delta\,,\nn\\
\hat\Gamma^\a{}_{\beta\mu} &=& -\ft12 b\, \Delta^{-1}\,
\delta^\a_\beta\, \del_\mu\, \Delta + \ft12 g^{\a\gamma}\, \del_\mu\,
g_{\beta\gamma}\,,\nn\\
\hat\Gamma^\mu{}_{\a\beta} &=& \ft12 b\, g_{\a\beta}\, \Delta^{-2}\, 
\del^\mu\, \Delta-\ft12 \Delta^{-1}\, \del^\mu\, g_{\a\beta}\,,\nn\\
\hat\Gamma^\a{}_{\beta\gamma} &=& \Gamma^\a{}_{\beta\gamma} - \ft12 b\,
\Delta^{-1}\, (\delta^\a_\gamma\, \del_\beta\, \Delta +
\delta^\a_\beta\, \del_\gamma\, \Delta - g_{\beta\gamma}\, \del^\a\,
\Delta) \,,
\eea
where
\be
\Gamma^\a{}_{\beta\gamma} \equiv \ft12 g^{\a\delta}\, (\del_\beta\,
g_{\delta\gamma} + \del_\gamma\, g_{\delta\beta} - \del_\delta\,
g_{\beta\gamma})=\Delta^{-1}\,
X_\a\,\mu_\a\,( \delta_{\beta\gamma} + \hat\mu_\beta\, \hat\mu_\gamma )
\,.
\ee
Note that $\del_\a$ means $\del/\del\mu_\a$, and that $\del^\a \equiv
g^{\a\beta}\, \del_\beta$.  
   
     We calculate the curvature using the expressions
\bea
\hat R^M{}_{NPQ} &=& \del_P\, \hat\Gamma^M{}_{NQ} - \del_Q\, \hat
\Gamma^M{}_{NP} + \hat\Gamma^M{}_{PR}\, \hat\Gamma^R{}_{QN} -
\hat\Gamma^M{}_{QR}\, \hat\Gamma^R{}_{PN}\,,\nn\\
\hat R_{NQ} &\equiv & \hat R^M{}_{NMQ} = 
\del_M\, \hat\Gamma^M{}_{NQ} - \del_Q\, \hat
\Gamma^M{}_{NM} + \hat\Gamma^M{}_{MR}\, \hat\Gamma^R{}_{QN} -
\hat\Gamma^M{}_{QR}\, \hat\Gamma^R{}_{MN}\,.
\eea
After some calculation, we find that
\bea
\hat R_{\mu\nu} &=& R_{\mu\nu} -\ft14 X_i^{-2}\, \del_\mu X_i\,
\del_\nu X_i + \ft12 \Delta^{-1}\, X_i^{-1}\, \mu_i^2\, \del_\mu X_i\,
\del_\nu X_i - \ft12 \Delta^{-2}\, \del_\mu\Delta\, \del_\nu\Delta
\nn\\
&&+\ft12 a\, (\Delta^{-2}\, \del_\lambda\Delta\, \del^\lambda\Delta -
\Delta^{-1}\, \square\, \Delta)\, g_{\mu\nu}\nn\\
&&-a\, \Big[ \sum_i X_i^2 - \Delta^{-1}\, X_i^2\, \mu_i^2\, \sum_j X_j
- 2\Delta^{-1}\, \, X_i^3\, \mu_i^2 + 2\Delta^{-2}\, (X_i^2\,
\mu_i^2)^2 \Big]\, g_{\mu\nu}\,,\\
\hat R_{\a\beta} &=& R_{\a\beta} +\ft12 b\, 
g_{\a\beta}\, \Delta^{-2}\, 
\square\, \Delta - \ft12 b\,  g_{\a\beta}\,\Delta^{-3}\,  
\del_\lambda\Delta\, \del^\lambda\Delta-\ft12 \Delta^{-1}\, \square\,
g_{\a\beta} \nn\\
&&+ \ft12 \Delta^{-1}\, g^{\gamma\delta}\, \del_\lambda
g_{\a\gamma}\, \del^\lambda g_{\beta\delta}
 -\ft14 \Delta^{-2}\, \del_\a\Delta\, \del_\beta\Delta - \ft12
\Delta^{-1}\, \nabla_\a\del_\beta\, \Delta \nn\\
&&- \ft14 b\, g_{\a\beta}\,\Delta^{-2}\,
\del_\gamma\Delta\, \del^\gamma \Delta+ \ft12 b\,
g_{\a\beta}\, \Delta^{-1}\, \nabla_\gamma\del^\gamma\, \Delta\,,\\
\hat R_{\a\mu} &=& - \ft12 \Delta^{-2}\, U\, ( X_\a^{-1}\, \del_\mu
X_\a - X_0^{-1}\, \del_\mu X_0)\, \mu_\a \,.\label{ralmu}
\eea
Note that in these expressions $\square$ means the Laplacian in the 
lower-dimensional spacetime, $\nabla_\a$ denotes the covariant derivative
with respect to the internal metric $g_{\a\beta}$, with its affine
connection $\Gamma^\gamma{}_{\a\beta}$, and $R_{\a\beta}$ is
the Ricci tensor calculated in this connection.

    Some useful {\it lemmata} which we used are
\bea
\del^\a\, \Delta\, \del_\a\, \Delta &=& 4 X_i^3\, \mu_i^2 - 4
\Delta^{-1}\, (X_i^2\, \mu_i^2)^2\,,\nn\\
\Gamma^\a{}_{\a\beta}  &=& \ft12 \Delta^{-1}\, \del_\beta\, \Delta 
+ \fft1{\mu_0^2}\, \mu_\beta\,,\nn\\
\nabla_\a\,\del^\a\, \Delta
&=& 2\sum_i X_i^2 - 2\Delta^{-1}\, X_i^2\, \mu_i^2\, \sum_j X_j + 4
\Delta^{-2}\, (X_i^2\, \mu_i^2)^2 \nn\\
&&- 4 \Delta^{-1}\, X_i^3\, \mu_i^2 +
\ft12 \Delta^{-1}\, \del^a\, \Delta\, \del_\a\, \Delta\,,\nn\\
R_{\a\b} &=& \Delta^{-1}\, \bar g_{\a\beta}\, \sum_\gamma X_\gamma\, 
-\Delta^{-2}\, (X_i^2\, \mu_i^2)\, \bar g_{\a\beta} \\
&&+ \Delta^{-2}\, (X_\a-X_0)(X_\beta -X_0)\, \mu_\a\, \mu_\beta -
\Delta^{-1}\, (X_\a-X_0)\, \delta_{\a\beta}\,,\nn\\
\square\, g_{\a\beta} &=& X_\a^{-3}\, \del_\lambda X_\a\, \del^\lambda
X^\a\, \delta_{\a\beta} + X_0^{-3}\, \del_\lambda X_0\, \del^\lambda
X_0\, \hat \mu_\a\, \hat \mu_\beta \nn\\
&&-4(X_\a\, \delta_{\a\beta} + X_0\, \hat\mu_\a\, \hat \mu_\beta) +
2 \bar g_{\a\beta}\, \sum_j X_j + \fft{4}{N}\, V\, g_{\a\beta}\,,\nn\\
g^{\gamma\delta}\, \del_\lambda g_{\a\gamma}\, \del^\lambda
g_{\beta\delta} &=& X_\a^{-3}\, \del_\lambda X_\a\, \del^\lambda
X_\a\, \delta_{\a\beta} + X_0^{-3}\, \del_\lambda X_0\, \del^\lambda
X_0\, \hat\mu_\a\, \hat \mu_\beta \nn\\
&& - \Delta^{-1}\, (X_\a^{-1}\, \del_\lambda X_\a - X_0^{-1}\,
\del_\lambda X_0)(X_\beta^{-1}\, \del^\lambda X_\beta - X_0^{-1}\,
\del^\lambda X_0)\, \mu_\a\, \mu_\beta\,.\nn
\eea
The quantities $\hat \mu_\a$
are defined by $\hat \mu_\a \equiv \mu_\a/\mu_0$, and the metric $\bar
g_{\a\beta}$ is defined by 
\be
\bar g_{\a\beta} \equiv  \delta_{\a\beta} + \hat\mu_\a\,
\hat\mu_\beta\,.
\ee
It is evident from (\ref{downmet}) that $\bar g_{\a\beta}$ is the
metric on the unit round $(N-1)$-sphere, corresponding to setting all
the $X_i=1$.  

\subsubsection{The Consistency of the Einstein Equation}

    With the results for the Ricci tensor from the previous section,
we can now verify that all components of the higher-dimensional 
Einstein equation are indeed consistently satisfied.  

   The higher-dimensional Einstein equation is
\be
\hat R_{MN} = \hat S_{MN}\,,\label{highereinst}
\ee
where 
\be
\hat S_{MN} = \fft1{2\, (D-1)!}\, \Big[ \hat F^2_{MN} - \fft{D-3}{D\,
(D-1)}\, \hat F^2\, \hat g_{MN} \Big]\,.
\ee

    The non-zero components of $\hat F_{M_1\cdots M_D}$ are given in
(\ref{fans}).  After some algebra, we find that
\bea
\hat F^2 &=& -D!\, \Delta^{-Da}\, ( U^2 + \Delta\, X_i^{-1}\,
\mu_i^2\, \del_\lambda X_i \, \del^\lambda X_i - \del_\lambda\Delta\,
\del^\lambda \Delta)\,,\nn\\
\hat F^2_{\mu\nu} &=& (D-1)!\, \Delta^{-2}\, \Big[
\Delta\, X_i^{-1}\, \mu_i^2\, \del_\mu X_i\, \del_\nu X_i-
\del_\mu\Delta\, \del_\nu\Delta \nn\\
&&\qquad \qquad\qquad - 
(\Delta\, X_i^{-1}\, \mu_i^2\, \del_\lambda X_i\, \del^\lambda X_i -
\del_\lambda\Delta\, \del^\lambda\Delta)\, g_{\mu\nu}
- U^2\, g_{\mu\nu}\Big] \,,\label{fsquared}\\
\hat F^2_{\a\beta} &=& -(D-1)!\, \Delta^{-2}\, (X_\a^{-1}\,
\del_\lambda X_\a - X_0^{-1}\, \del_\lambda X_0)(
X_\b^{-1}\, \del^\lambda X_\b - X_0^{-1}\, \del^\lambda X_0) \,
\mu_\a\, \mu_\b\, \,,\nn
\eea
where, as usual, $U$ is given by
\be
U = 2 X_i^2\, \mu_i^2 - \Delta\, \sum_i X_i\,.
\ee

   Thus we find that $\hat S_{MN}$ is given by
\bea
\hat S_{\mu\nu} &=& \ft12 \Delta^{-1}\, X_i^{-1}\, \mu_i^2\, \del_\mu
X_i\, \del_\nu X_i - \ft12 \Delta^{-2}\, \del_\mu\Delta\,
\del_\nu\Delta \nn\\
&& -\fft1{D-1}\, \Delta^{-2}\, (U^2 - \del_\lambda\Delta\,
\del^\lambda\Delta + \Delta\, X_i^{-1}\, \mu_i^2\, \del_\lambda X_i\,
\del^\lambda X_i)\, g_{\mu\nu}\,,\label{smunu}\\
\hat S_{\a\beta} &=& \ft12 b\, \Delta^{-3}\, U^2\, g_{\a\beta} +
\ft12 b\,\Delta^{-2}\, g_{\a\beta}\, X_i^{-1}\, \mu_i^2\, \del_\lambda
X_i\, \del^\lambda X_i -\ft12 b\, \Delta^{-3}\, \del_\lambda \Delta\,
\del^\lambda \Delta\, g_{\a\b}\nn\\
&&-\ft12 \Delta^{-2}\,  (X_\a^{-1}\, \del_\lambda X_\a - X_0^{-1}\,
\del_\lambda X_0)(X_\beta^{-1}\, \del^\lambda X_\beta - X_0^{-1}\,
\del^\lambda X_0)\, \mu_\a\, \mu_\beta\,,\label{salbe}\\
\hat S_{\a\mu} &=&- \ft12 \Delta^{-2}\, U\, ( X_\a^{-1}\, \del_\mu
X_\a - X_0^{-1}\, \del_\mu X_0)\, \mu_\a\,.\label{salmu}
\eea

    To verify that the components 
$\hat R_{\mu\nu}=\hat S_{\mu\nu}$ of the higher-dimensional Einstein
equation indeed imply the lower-dimensional Einstein equation 
(\ref{ddeinst}), we simply need to substitute the above results into
(\ref{highereinst}).  It is also necessary to use the scalar equations
of motion in (\ref{xeq0}), from which we can deduce that
\be
\square\, \Delta = X_i^{-1}\, \mu_i^2\, \del_\lambda X_i\,
\del^\lambda X_i + 4 X_i^3\, \mu_i^2 - 2 X_i^2\, \mu_i^2\, \sum_j X_j
- \fft4{N}\, \Delta\, V\,.
\ee

   Putting all the results together, we find that indeed all the
$\mu_i$ dependence cancels out in the $\hat R_{\mu\nu}=\hat
S_{\mu\nu}$ equation, and we correctly reproduce the lower-dimensional
Einstein equation in (\ref{ddeinst}).

        After some algebra, using the {\it lemmata} given previously, we find
that the components $\hat R_{\a\beta}$ of the Ricci tensor of the
higher-dimensional metric are simply given by
\bea
\hat R_{\a\beta} &=& \ft12 b\, \Delta^{-3}\, U^2\, g_{\a\beta} +
\ft12 \Delta^{-2}\, g_{\a\beta}\, X_i^{-1}\, \mu_i^2\, \del_\lambda
X_i\, \del^\lambda X_i -\ft12 b\, \Delta^{-3}\, \del_\lambda \Delta\, 
\del^\lambda \Delta\nn\\
&&-\ft12 \Delta^{-1}\,  (X_\a^{-1}\, \del_\lambda X_\a - X_0^{-1}\,
\del_\lambda X_0)(X_\beta^{-1}\, \del^\lambda X_\beta - X_0^{-1}\,
\del^\lambda X_0)\, \mu_\a\, \mu_\beta\,.
\eea
Note that we have made use of the equations of motion for the $X_i$
fields in simplifying this expression.  It is now straightforward to
see that this is exactly equal to the expression for $\hat
S_{\a\beta}$ obtained in (\ref{salbe}).  Finally, the components $\hat
S_{\a\mu}$ given in (\ref{salmu}) agree precisely with the
corresponding components $\hat R_{\a\mu}$ found in (\ref{ralmu}).  Thus
the consistency of the reduction Ansatz is completely verified.

\section{Scalar Potentials in $D=3$}

       In the previous sections, we proved the consistency of the
embedding of the diagonal symmetric potentials in the relevant higher
dimensions.  The number of scalars $N$ and the (lower) dimension $D$
are related by (\ref{ndrel}).  As was shown in \cite{dist}, the various
$D$-dimensional multi-charge extremal AdS domain walls supported by
these scalars can be oxidised back to solutions of eleven-dimensional
supergravity ($D=4$ and $D=7$) or type IIB supergravity $D=5$).  These
higher-dimensional solutions correspond to ellipsoidal continuous
distributions of M5-branes, M2-branes and D3-branes respectively
\cite{klt,fgpw,BS,dist,BS2,BBS}.

   For general values of $D$ the relation (\ref{ndrel}) would imply a
non-integral value for $N$, and no consistent embedding exists.  The
relation becomes singular for the case $D=3$.  Thus contrary to what
one might have hoped, the pattern of consistent embeddings does not
seem to extend to an $S^3$ reduction from $D=6$ to $D=3$.  Indeed, it is
straightforward to show that the ellipsoidal continuous distributions
of dyonic strings that exist in $D=6$ do not lend themselves to
consistent reductions to $D=3$.

   In this section, we discuss an alternative reduction to a gauged
$D=3$ supergravity,  in which there is a massive scalar field.  
The three-dimensional bosonic Lagrangian is given by
\be
e^{-1}\, {\cal L}_3 = R -\ft12(\del\phi)^2 -\ft12 g^2\,
\Big( \fft{1}{a_1^2} e^{a_1\phi} -\fft{1}{a_1a_2} e^{a_2\phi} \Big)\,,
\label{d3lag1}
\ee
where $a_1^2 =4/k + 4$ and $a_2=4/a_1$.  The integer $k$ can take the
values 1, 2, or 3.  The values $k=2$ and $k=3$ correspond to the $S^3$
reduction of $D=6$ simple (chiral) supergravity and the $S^2$ reduction of
$D=5$ simple supergravity respectively, and $\phi$ is the associated
massive breathing mode \cite{instanton}.

     The case of $k=1$ corresponds to the $S^1$ Scherk-Schwarz
reduction of the Freedman-Schwarz model.  To show this, we begin from
the Lagrangian for the gravity plus scalar sector of the $D=4$
Freedman-Schwarz model \cite{FS}, which can be obtained as a singular
limit \cite{d4gauge} of the $N=4$, $D=4$, $SO(4)$ gauged supergravity
\cite{dfr}:
\be
\hat e^{-1} {\cal L}_4 = \hat R - 
\ft12 (\del\hat \phi)^2 -\ft12 (\del\chi)^2\,
e^{2\hat \phi} +\fft12 g^2 e^{\hat \phi}\,.
\ee
Dimensionally reducing this theory on a coordinate $z$, where the
axion $\chi$ is allowed to take the generalised Scherk-Schwarz form 
$\chi = m\, z$, we obtain the three-dimensional
scalar Lagrangian
\be
e^{-1} {\cal L}_3 = R -\ft12 (\del\hat\phi)^2 -\ft12(\del\varphi)^2
-\ft12 m^2\, e^{2(\hat\phi +\varphi)} +\ft12 g^2\, e^{\hat\phi
+\varphi}\,.\label{d3lag2}
\ee
Since the original dilaton $\hat\phi$ and the dilaton $\varphi$ coming
from the dimensional reduction occur everywhere in the same
combination, we see that it is consistent to truncate out the
combination $\hat\phi -\varphi$.  Making the redefinition $\phi\equiv
(\hat\phi + \varphi)/\sqrt2$, the Lagrangian (\ref{d3lag2}) reduces to
(\ref{d3lag1}) with $k=1$.  The three Lagrangians in (\ref{d3lag1})
all give rise to supersymmetric domain-wall solutions in $D=3$
\cite{instanton,clpwall}.

\section{Conclusion}

     In this paper, we have provided a complete proof of the
consistency of the Kaluza-Klein reduction Ans\"atze that were
presented in \cite{dist}, which describe the embedding of certain
$N$-scalar truncations of the maximal gauged supergravities in $D=4$,
7 and 5, {\it via} spherical reductions on $S^7$, $S^4$ and $S^5$
respectively.  The $N$ scalars, with $N=8$, 5 and 6 respectively,
correspond to the diagonal elements in the $SL(N,R)/SO(N)$
submanifolds of the full scalar manifolds in the corresponding maximal
supergravities.  (Actually, there are really only $N-1$ independent
scalars in these truncations, on account of a unit-determinant
condition on the scalars in the coset.)  Our proof included a complete
verification of the consistency of the reduction of the
higher-dimensional Einstein equation, which is usually the most
calculationally difficult part of the procedure.  

     For $D=7$, our results are consistent with the full Kaluza-Klein
$S^4$ reduction that was recently obtained explicitly in \cite{nvv0,nvv}.
For $D=4$, they are compatible with the implicit proof of the
consistency of the complete $S^7$ reduction, presented in
\cite{deWitnicolai}.  Furthermore, our results provide a complete
proof of the validity of the explicit expressions presented in
\cite{dist} for the Ans\"atze for the eleven-dimensional fields,
which, especially in the case of the 4-form field strength, are not
straightforward to extract from the results presented in
\cite{deWitnicolai}.

    Finally, in $D=5$ our results provide further evidence for the
conjectured consistency of the $S^5$ reduction of type IIB supergravity
to give maximal $SO(6)$ gauged supergravity in $D=5$.  

    We also considered the special case of scalar theories in $D=3$
that arise from dimensional reduction.  This dimension lies outside
the set of cases covered by the previous discussion, on account of a
degeneration in the formula (\ref{ndrel}) relating the dimension to
the number of scalar fields.  Instead, we described the set of three
theories (\ref{d3lag1}) arising as the scalar sectors of sphere
reductions from $D=6$, $D=5$ and $D=4$.  In the case of $D=4$, we
showed how the single-scalar Lagrangian (\ref{d3lag1}) arises from a
Scherk-Schwarz $S^1$ reduction of the $D=4$ Freedman-Schwarz model,
accompanied by a further consistent truncation of one combination of
the two resulting dilatonic scalar fields.

\section*{Acknowledgements}
 
    We are grateful to Jim Liu and Tuan Tran for helpful discussions.

\end{document}